# A fiber based diamond RF B-field sensor and characterization of a small helical antenna

M. M. Dong, Z. Z. Hu, Y. Liu, B. Yang, Y. J. Wang, G. X. Du*,

College of Telecommunication & Information Engineering, Nanjing University of Posts and Telecommunications, No. 66 Xin Mofan Road, Nanjing, China.

We present a microwave B-field scanning imaging technique using diamond micro-crystal containing nitrogen vacancy center that is attached to a fiber tip. We propose a pulsed modulation technique, enabling the implementation of a variety of pulsed quantum algorithm for state manipulation and fast readout of spin state. A detailed mapping of the magnetic B-field distribution of a helical antenna with sub-100 micron resolution is presented and compared with numerical simulations. This fiber based microwave B-field probe has the advantage of minimized invasiveness, small overall size, will boost broad interest in a variety of applications where near field distribution is essential to device characterization, to name a few, antenna radiation profiling, monolithic microwave integrated circuits failure diagnosis, electromagnetic compatibility test of microwave integrated circuits and microwave cavity field mode mapping.

*Corresponding author: G. X. Du

Email: duguanxiang@njupt.edu.cn

Developing high sensitive and high resolution imaging techniques of magnetic field with frequency from DC to microwave has been at the forefront of research. Novel techniques has enabled a range of application and discoveries, such as magnetic resonance imaging [1 – 2], sensing of a single electron and nuclear spin [3], imaging of a superconductor vortex [4], and imaging topological structures in magnetic thin film [5]. On the macroscale, geomagnetic survey and space magnetometers require ultrahigh field sensitivity and accuracy, which is enabled by Alkali vapor cell magnetometer [6,7]. Sensitivity of sub-pT/ $\sqrt{Hz}$ is now commercially available. Magnetic signal from the human neural activity is on the order of femtoTesla [8]. Such a weak field strength is detected dominantly by SQUID [9] in functional magnetic resonance imaging and is believed to be replaced in the near future by Alkali vapor cell magnetometer [10]. In other applications, however, knowledge on the spatially resolved magnetic field distribution is indispensable, where magnetic sensor is placed in the close proximity of the tiny magnetic structures. For example, magnetic bit of ~10 nanometer width is resolved by a sensitive spintronic read head in the state-of-the-art hard disk drive [11]. Nuclear magnetic resonance at the molecule or cell level is only possible very recently with the development of diamond quantum sensor [12-14]. Nanometer resolved imaging of microwave field has been enabled by a single nitrogen vacancy (NV) center in diamond [15, 16]. In this work, we focus on developing field sensing technique for applications where spatial resolution at micrometer scale is desired [17-20]. Special attentions are paid on robust operation and experimentally easy implementation for practical applications. We demonstrate a high-resolution microwave B-field scanning system using a diamond RF field probe that contains NV centers. Sub-100 micron diamond crystal fixed to the tip of a tapered fiber is used as a magnetic field sensor, which is low cost and robust at continuous operation [21, 22]. This work first describes the characteristics of the diamond B-field sensor, the principle of pulsed modulation technique and then a full characterization of the field distribution around and inside the helical antenna is presented.

The nitrogen vacancy (NV) color center consists of a substitutional nitrogen atom and an adjacent vacancy, which is a photoluminescent point defect in diamond crystal [23-25]. The negatively charged NV center has long spin coherence time at ambient condition and can be prepared and readout optically. The ground state of NV center has a zero-field splitting of 2.87 GHz between the magnetic sub-levels $m_s= 0$ and $m_s= \pm 1$ at room temperature due to spin-spin interaction. When an external magnetic field $B_0$ is applied, the degeneracy of the spin sublevels $m_s= \pm 1$ is lifted by Zeeman effect. By adjusting the relative orientation of the external magnetic field and the four crystal NV axes, totally eight microwave dipole transitions in the ground state can be observed by the well-developed optically detected magnetic resonance (ODMR) techniques. The transition between $m_s= 0$ and $m_s= +1$ or $m_s= -1$ is magnetic dipole allowed. They form an idea two-level-atom quantum system. Resonant microwave B-field drives the spin in a closed Rabi cycle on a Bloch sphere. The oscillation Rabi frequency is a quantitative measure of the amplitude of field component transverse to NV axis.

For field mapping applications, however, running a full Rabi sequence at each position is very time consuming and not practical for real world applications. Instead of running a full Rabi sequence to extract the oscillation frequency at each pixel, we performed a pulsed Rabi sequence with a single fixed and rather short duration of microwave pulse [15, 19]. To observe small modulation in the fluorescence signal due to short microwave interaction time, a new modulation scheme is indispensable, where the amplitude or frequency of microwave is modulated, which results in a sideband signal of pulsed fluorescence. This can be formulated as follows.

Microwave B-field modulates the fluorescence and observes a decayed sinusoid behavior as [15]

$$I = I_0 - i_0 \sin(2\pi\gamma B_{mw}t)\exp(-t/\tau) \quad (1)$$

where total FL intensity is a constant $I_0$ superimposed by a microwave modulated component $i_0$. $\gamma$ is the gyromagnetic ratio of electron spin, $t$ is the pulse duration of microwave and $\tau$ is the Rabi dephasing time due to limited spin life time, inhomogeneity in both static and microwave field.

In the limit of $\gamma B_{mw}t \ll 1$ (linear regime), Eq. (1) can be approximated by

$$I = I_0 - i_0(2\pi\gamma B_{mw}t)\exp(-t/\tau) \quad (2)$$

The modulation depth, defined as $i_0/I_0$, is ~ 30% in single NV magnetometry, but significantly decreased in NV ensemble, due to the fact that only one in eight spin transitions is active at resonance and coupling to nitrogen nuclear spin [26]. The other transitions contribute a large FL background and reduces the modulation depth to a maximum of 4%. This is then further reduced by a factor of three, to ~1%, due to the hyperfine coupling with nitrogen nuclear spin ($I=1$). Higher modulation depth can be achieved by inducing preferential growth along certain crystal axis in NV center manufacture process [27].

Commercial available laser intensity is usually stabilized on the order of 0.5%, comparable with the modulation depth. To detect even smaller modulation component in a large background $I_0$, amplitude modulation of microwave B-field and lock-in technique is adopted in conventional CW ODMR experiments. In the pulsed ODMR scheme, lock-in technique is not straightforward, but is still applicable with necessary modification to the detection scheme. In addition to apply laser synchronized pulse to microwave, microwave is further amplitude modulated at the source output with a relatively low modulation frequency $f_m$, usually at kHz. This modulated B-field generates a sideband signal of the fluorescence pulse, with a carrier frequency of $f_c$. Demodulation at the sideband frequency provides a significantly improved signal to noise ratio (SNR) against the intensity fluctuation of excitation laser, a minimum measurable modulation depth of 20 ppm is observed with resolution bandwidth of 3 Hz. Taking into account of the maximum modulation depth of 1%, the highest SNR of 500 can be observed and is verified in our experiments. This is also the field measurement dynamic range which is important, corresponding to a dynamic range of 54 dB in microwave power.

In a typical Rabi sequence, which consists of staggered laser and microwave pulses, the laser pulses polarize and detect the ground spin state, the microwave pulses drive the spin state into Rabi cycles between spin sub-levels. The FL is thus a pulsed signal, resonant microwave driving results in a reduced FL level. When microwave B-field is amplitude modulated, a sideband signal centered at the FL carrier frequency with a frequency shift same as the modulation frequency is observed. Noted that this pulsed lock-in technique is a complementary to the conventional CW lock-in technique, where the fluorescence is a DC value and lock-in amplifier measures signal amplitude at the modulation frequency. Indeed, CW double resonance technique is not suitable for microwave field measurement, where polarization by laser and depolarization by microwave exist simultaneously, and thus the modulation depth is no longer a meaningful quantity to evaluate the field strength, which is dependent on the laser intensity.

The sketch of the fiber based diamond microwave B-field sensing and scanning system is shown in Fig. 1. All optical components and the device under test (DUT) are arranged on the vibration isolation optical table. For micrometer scanning imaging, environmental vibration should be carefully minimized. Spectrometer, microwave signal generator, signal analyzer and power supplies are placed on an elevated platform above the optical table. The 532nm laser reflected from a dichroic mirror (DM), is focused onto the flat end of a tapered fiber by an objective lens, the transmitted light is coupled to the diamond crystal which is fixed at the tip of the fiber to excite the NV color center. Fluorescence signal is collected by the same fiber and objective lens, transmitted through DM and focused to the active area of an avalanche photodetector (APD). In front of the APD, long-pass filter (660 – 800 nm, covering the red fluorescence spectra of NV center) is inserted to block the green light. Signal analyzer performs spectral analysis of output from APD. Microwave signal generator (SSG-6400HS, Mini-Circuits Inc.) is pulsed by a PIN switch with a typical rise/off time of a few nanosecond, and is power amplified before connected to the DUT. Microwave isolator is used to protect the power amplifier from damage due to large power reflection from DUT. The DUT, a helical antenna in this work, is mounted on the motorized XY stage (SC300, Zolix Instruments Co., Ltd.) for mapping of field distribution. The diamond fiber probe is fixed on the optical table and brought to the near surface of DUT. A permanent magnet applies a small static magnetic field to split up the degeneracy of $m_s = \pm 1$ state.

Diamond microcrystal is commercially available, which is an ion implanted high pressure high temperature (HPHT) diamond crystal and milled to around 60 micron in size [28]. Smaller size gives rise to higher spatial resolution but at the price of reduced fluorescence intensity. A compromise between spatial resolution and FL intensity is found in this work. Tapered fiber is prepared by heating the fiber in the flame of a lighter and stretching it to form desired taper geometry. A typical tapered fiber has a fiber end diameter of 70 micrometer which is tapered over a distance of 5 millimeter from a fiber core diameter of 200 micron. Simple ray optics suggests that taper geometry enhances the numerical aperture of fiber, which enhances both the collection efficiency and excitation laser intensity. However, detailed discussion is out of the scope of this work. Transfer of a sub-100 micron crystal onto the fiber tip is under a 500 times digital microscope and glued with UV curable glue. Detection of weak fluorescence intensity is enabled by high responsivity APD without further electronic signal amplification. To implement pulsed Rabi sequence, the degeneracy of $m_s = \pm 1$ should be lifted and totally eight peaks should be fully resolved in the ODMR spectra, which is done manually by positioning a permanent magnet around the diamond crystal. The magnet is mounted on a manual stage for smooth adjustment of the magnetic field strength and orientation at the position of diamond crystal.

Pulsed ODMR sequence consists of alternative laser and microwave pulses. The laser pulse (ON/OFF = 500/500 nanosecond) is optimized for efficient polarization as well as readout of the ground spin state. The microwave pulse length is set to 50 nanosecond within laser off time window. Microwave is amplitude modulated at the source, with a relatively low frequency of 1 kHz. In frequency domain, the pulsed fluorescence thus has a frequency component ($V_c$) at 1 MHz and the modulation results in a sideband signal ($V_s$) of the fluorescence carrier signal. In linear regime ($\gamma B_{mw} t \ll 1$), sideband signal scales linearly with the carrier signal, the ratio of them, $V_s/V_c$, is defined as modulation depth. We demonstrate that the minimum resolvable modulation depth, limited by the noise floor of the whole system, is ~20 ppm with resolution bandwidth (RBW) of 3 Hz. Reducing RBW further improves the minimum detectable microwave B-field but at the cost of longer measurement time. At the weak field limit, having a typical Rabi Pi pulse length of 4 microsecond (Bmw~0.0446 Gauss), the modulation depth has a maximum of ~1%, the minimum resolvable field is 0.89e-4 Gauss. Taking RBW into account, the sensitivity is thus 0.5e-4 Gauss/ √ Hz, or 5 nT/ √ Hz. Fig. 2 (a) shows well resolved pulsed ODMR spectra. The solid line is the fitting data using a combination of eight Lorentzian lines with linewidth of ~6 MHz.

Pulsed Rabi sequence is similar to the ODMR sequence, with microwave pulse length scanned within laser off time window. The microwave B-field strength $B_{mw}$ is then deduced from the measurement of Rabi frequency $\Omega_R$ through $\gamma B_{mw} = \hbar \Omega_R$, where $\gamma$ and $\hbar$ are electron gyromagnetic ratio and the reduced Planck constant, respectively. Despite the relatively short spin relaxation time of the dense HPHT sample, several Rabi cycles are observed for microwave pulse duration up to ~4 microsecond. A typical Rabi oscillation curve is shown in Fig. 2 (b). The solid line is a fit with a decayed sinusoid function. Fit to the fast oscillation in the short microwave pulse length region is poor, which is due to microwave field inhomogeneity. A fast Fourier transform of the Rabi curve is shown in the inset.

We demonstrate the field mapping both inside and around a helical antenna. The antenna is a twenty and a half turn copper solenoid, 33 mm long and 5 mm in diameter, widely used in telecommunications [20, 29]. Knowledge on the field distribution of the antenna is essential to its directivity. To the best of our knowledge, however, very few are reported about the actual field distribution [30], either inside or outside the antenna. This might be due to the bulky size and invasiveness of conventional B-field probe. The diamond micro-crystal of sub-100 micrometer size, attached to an optical fiber, can be conveniently brought into the interior of the antenna, with minimized invasiveness. The field mapping is recorded by scanning the antenna in a defined 2D plane, with a fixed short microwave pulse length in the linear regime. The field strength is then linearly proportional to the side band signal amplitude. The pulse length is predetermined in a separate rough scan near antenna surface, a full Rabi sequence is subsequently performed at the strong field location. This procedure ensures that the microwave pulse length across the whole image is in the linear regime. Noted that the high sensitivity detection of modulation depth (down to ~20 ppm) promises the setting of a rather short microwave interaction time. Fig. 3 shows a line scan of the field distribution along the center axis of the helix, measured at 2.857 GHz. Standing-wave-like feature is observed, which is a result of interference of propagating and counter propagating microwave reflected at the end of the helix. Fringe contrast of unity indicates large reflection coefficient, associated with the 50 Ohm impedance mismatch with the signal source. The measured field distribution is compared with the numerical simulation by HFSS software. Simulation uses the actual geometry parameters of the antenna. We also performed a 2D field mapping outside the antenna in the radial plane, which is shown in Fig. 4. The mapping plane is illustrated. A line scan of both the simulation and measurement data is shown in illustration 2,3. The combined high spatial resolution and field sensitivity enables detailed knowledge about the near field pattern of the antenna. Characteristic field lobes are observed, associated with the current density distribution along the antenna conductor. Good consistency between the simulation and measurement results is observed in both cases. Noted that the simulation is performed without introducing any fitting parameters.

In this work, we present a fiber based diamond sensor for microwave B-field with micrometer resolution. Pulsed lock-in technique is proposed in the scheme of Rabi cycle. Resonant microwave driving generates a sideband signal on the excitation laser pulse, the amplitude of which is linearly proportional to the B-field strength in the limit of ultrashort microwave pulse duration. Absolute field strength mapping is done by running an additional full Rabi sequence. This technique introduces minimized invasiveness to the field under test, is robust at operation and has small overall size, could boost a range of real world applications where micrometer resolution image of B-field distribution is essential to device characterization.

Acknowledgement

G. X. Du acknowledge financial support from Jiangsu distinguished professor program (Grant No. RK002STP15001) and NJUPT principal distinguished professor program (Grant No. NY214136).

**Figure captions**

Fig. 1. (a) Sketch of the microwave B-field scanning imaging setup. RF field probe is a micron-sized diamond crystal, fixed at a tapered fiber tip. (b) Pulsed lock-in technique control sequence. Microwave is amplitude modulated by MW_AM pulse at the signal source and further gated by MW_S synchronized with laser control pulse.

Fig. 2. (a) Typical pulsed lock-in ODMR curve showing eight resonance peaks. (b) Typical Rabi oscillations observed in NV ensemble. A fast Fourier transform of the Rabi curve is shown in (c).

Fig3. Axial field distribution inside a helical antenna and HFSS simulation results. Upper panel shows the field distribution inside the helical antenna.

Fig. 4. Microwave B-field distribution outside the helical antenna in the axial plane (lower) and HFSS simulation results (upper). Illustration 1 shows the relative position

of the field mapping plane with respect to the helical antenna. Illustration 2 shows a field line scan along the upper bound of the field map. Illustration 3 shows a field line scan along radial direction of the field map.

**Figures**

Fig.1

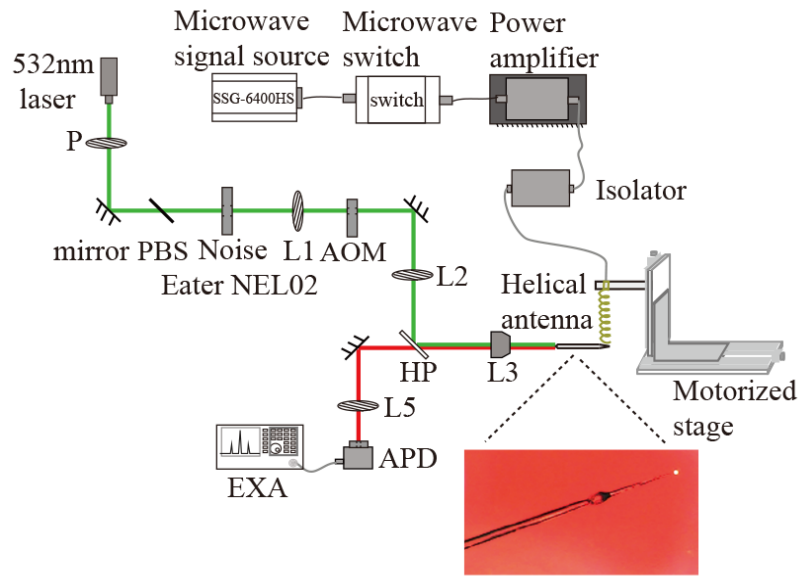

(a)

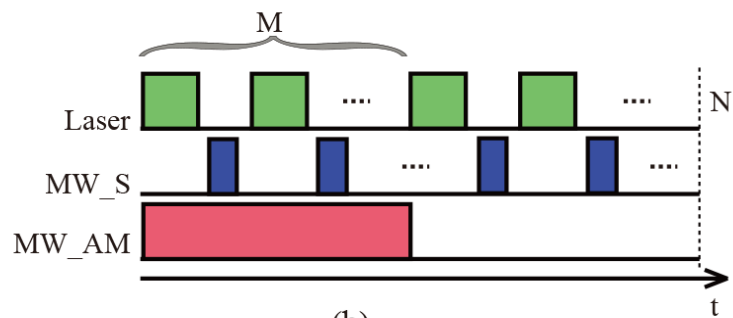

(b)

Fig. 2

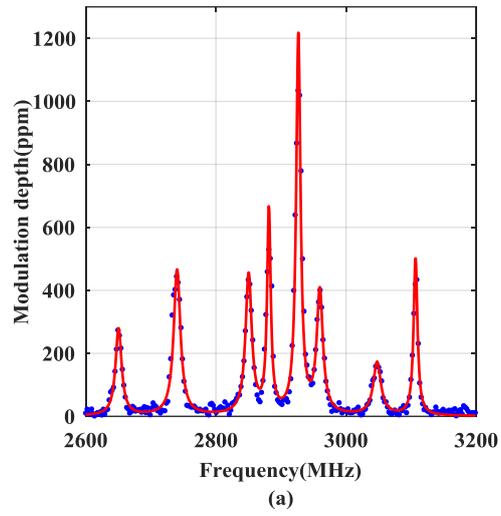

(a)

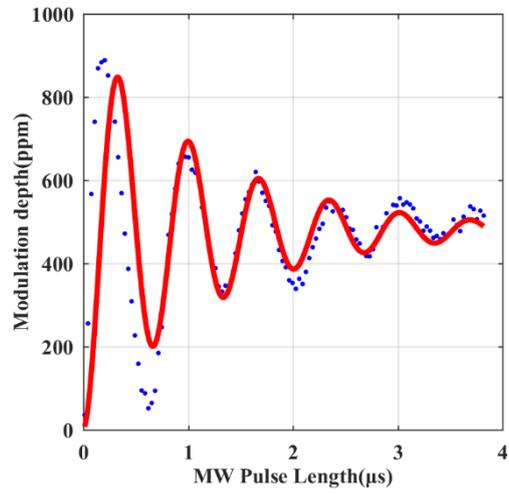

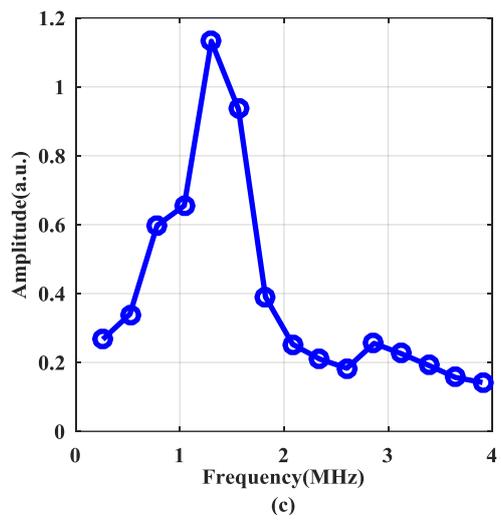

(c)

Fig.3

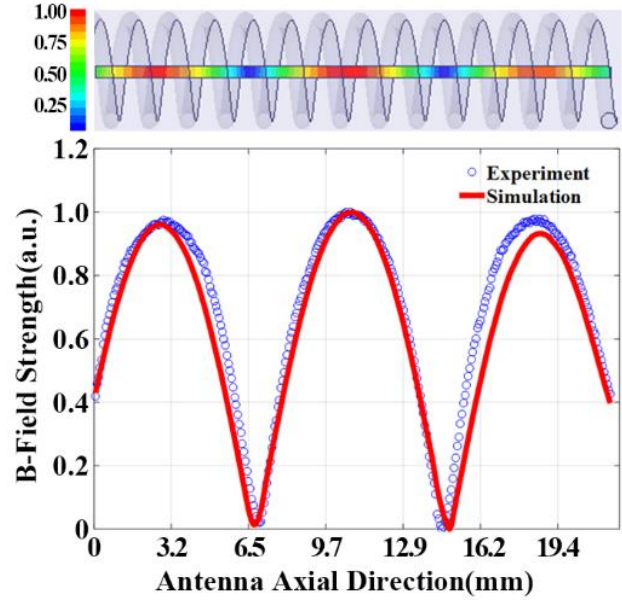

Fig.4

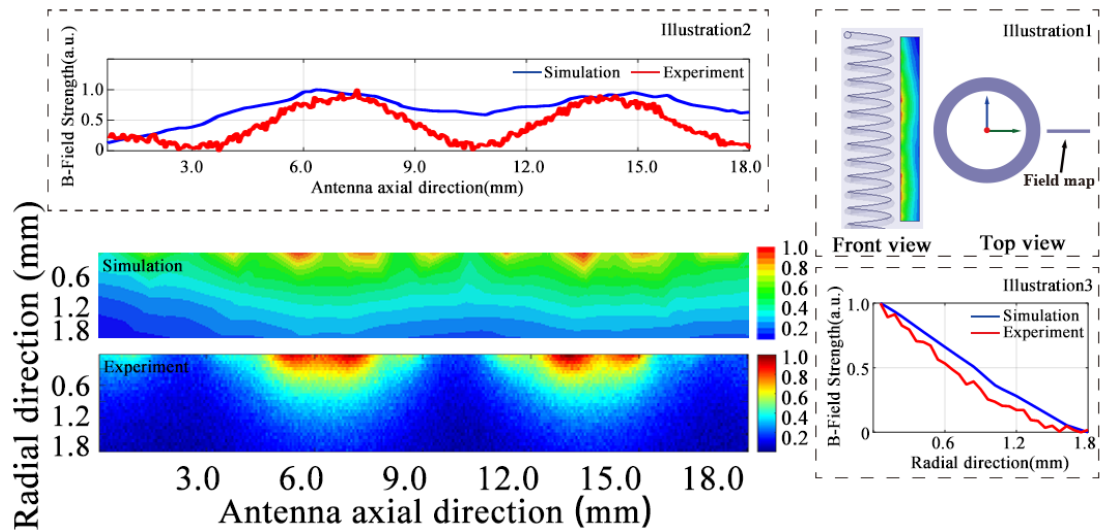